\documentclass{PoS}
\title{Joint analysis of TeV blazar light curves with FACT and HAWC}

\ShortTitle{Joint blazar analysis with FACT and HAWC}

\author{\speaker{Daniela Dorner}, for the FACT Collaboration
   \thanks{Complete FACT author list at http://fact-project.org/collaboration/icrc2017\_authorlist.html }\\
        Universit\"at W\"urzburg, Institute for Theoretical Physics and Astrophysics,
Emil-Fischer-Str.\ 31, 97074 W\"urzburg,Germany\\
        E-mail: \email{dorner@astro.uni-wuerzburg.de}}

\author{Robert Lauer, for the HAWC Collaboration\thanks{Complete HAWC author list at 
http://www.hawc-observatory.org/collaboration/icrc2017.php}\\
        University of New Mexico, Department of Physics and Astronomy, 
Albuquerque, NM, USA\\
        E-mail: \email{rjlauer@unm.edu}}

\abstract{Probing the high energy emission processes of blazars through their variability relies crucially on long-term monitoring. We present unprecedented light curves from unbiased observations of very high energy fluxes from the blazars Mrk\,421 and Mrk\,501 based on a joint analysis of data from the First G-APD Cherenkov Telescope (FACT) and the High Altitude Water Cherenkov (HAWC) Observatory. Thanks to an offset of 5.3 hours of the geographic locations, a complementary coverage of up to 12 hours of observation per day allows us to track variability on time scales of hours to days in more detail than with single-instrument analyses. Complementary features, such as better sensitivity thanks to a lower energy threshold with FACT and more regular coverage throughout the year with HAWC, provide valuable cross checks and extensions to the individual analyses. Daily flux comparisons for both Mrk\,421 and Mrk\,501 show largely correlated variations with a few significant exceptions. These deviations between measurements can be explained through fast variability within a few hours and will be discussed in detail.}

\FullConference{35th International Cosmic Ray Conference -- ICRC217-\\
		10-20 July, 2017\\
		Bexco, Busan, Korea}

\begin{document}

\section{Introduction}


The First G-APD Cherenkov Telescope (FACT) is an imaging air Cherenkov telescope (IACT) with 9.5\,sqm mirror area. It is located on the Canary Island of La Palma at 2200\,m a.s.l. Apart from long-term monitoring of bright TeV blazars, one major goal was the proof of principle of silicon based photosensors (SiPM, aka Geiger-Mode Avalanche Photo Diodes, G-APDs). These novel photo-sensors show an excellent performance \cite{2014JInst...9P0012B} which makes them ideal for long-term monitoring. As SiPMs do not degrade when exposed to bright ambient light, not only the duty cycle of the instrument is enlarged but also the gaps in the light curves due to full moon are minimized \cite{FACTHighlightsICRC2017}. 


The HAWC Observatory is a second generation water Cherenkov gamma-ray 
observatory, sensitive to primary energies between approximately 500 GeV and 100 TeV.
The HAWC Observatory is located at an elevation of 4,100~m a.s.l.\ on the flank of the Sierra Negra volcano in Puebla, Mexico. It has been operating since March 2015 as an array of 300 detectors, housed in large water tanks, that instrument an area of 22,000~m$^2$. HAWC's main distinctions compared to other currently operating very high energy (VHE) gamma-ray detectors are its large duty cycle, $\sim95$~\%, and its wide field-of-view of $\sim2$~steradians. HAWC is monitoring every source that transits through a $90^{\circ}$ cone centered on local zenith for up to 6 hours per day. An overview 
of the analysis is provided in~\cite{HawcCrab}, a catalog of steady sources seen in the first 17 
months in \cite{HawcCatalog2016}, and first results from daily monitoring in~\cite{HawcLightcurves}. 


While FACT has to point to the source and therefore can observe only a limited sample of sources, HAWC is continuously monitoring two thirds of the sky. Due to the detection technique HAWC is furthermore independent of the sun and the weather and provides a more continuous light curve. FACT on the other hand provides more sensitive measurements thanks to the lower energy threshold and better angular resolution. Being offset by 5.3 hours in the location, combining the data of the two instruments provides a wider coverage in time of up to 12~hours. With the joint analysis laid out in these proceedings, we aim to use these complementary features to investigate the time structure as well as spectral variability of VHE flares observed from the blazars Mrk\,421 and Mrk\,501. The goal is to better understand and model acceleration mechanisms in these sources.

\section{FACT Analysis}

For the FACT data, nightly binning and 20 minute binning has been used based on the results of the quick look analysis as described in \cite{2015arXiv150202582D}. In addition a data selection has been performed rejecting data affected by bad weather according to the method discussed in \cite{2013arXiv1311.0478D}, including an optimization of the cut parameters for time periods with different performance. Based on data from the Crab Nebula as a TeV-energy calibration source, the dependence of the measured excess rates from the zenith distance has been studied to determine correction factors for the excess rate as described in \cite{MMahlkeICRC2017}. The correction factor for the dependence of the excess rate on the ambient light has been determined in a similar way. For each time period of different performance of the system (including the detector and the analysis), the conversion factor from excess rate to Crab Units has been determined. Using this and the integral flux of the Crab Nebula above 750\,GeV, excess rates are converted to fluxes. 


\section{HAWC Analysis}

\subsection{Light Curves}

The HAWC data included in the analysis presented here are the daily flux light curves for Mrk\,421 and Mrk\,501, spanning November 2014 to April 2016, as recently published in~\cite{HawcLightcurves}. 
Each daily measurement corresponds to the average flux during one full transit over HAWC of approximately 6 hours, with 90\% of the sensitivity stemming from the central 4 hours where 
zenith distances for observation are smallest. While it is possible to achieve resolution of shorter time scales with HAWC during periods of high flux, the increasing statistical uncertainties are on average too large to allow shorter intervals for the analysis discussed here.

Daily flux measurements with HAWC rely on determining a spectral model (see next section) and then 
fitting only the normalization of this model with the data taken during each transit via a maximum 
likelihood approach. 
These normalization values are used to calculate daily photon 
fluxes by integrating the differential flux parametrization above an energy threshold. 
In a next step, the Bayesian blocks algorithm \cite{Scargle2012} was used to find a segmentation of the light curves into different flux states, choosing a prior so that the false positive 
probability for wrongly identifying a change of flux where none occurred is 5\% per light curve.

\subsection{Spectral Modeling}

In~\cite{HawcLightcurves}, the 
spectral model for both Mrk 421 and 501 was a power law with an exponential cut-off, based on a 
fit to the combined data from 17 months. For the work presented here, we improve on this in 
two ways:

\begin{enumerate}
\item We include extra-galactic background light (EBL) in the spectral modeling by using the 
parametrization from \cite{Gilmore2012} to attenuate an intrinsic model flux as a function of 
energy and distance, with redshifts z=0.031 for Mrk 421 and z=0.034 for Mrk 501.  We can thus distinguish a high-energy cut-off in the observations due to the 
EBL from intrinsic spectral curvature. 

\item Instead of only using a fixed average spectral model for the whole period, we fit the power 
law index of the intrinsic model for each period that constitutes a different flux state. This can 
trace the variability of spectral features in the sources.
\end{enumerate}

The short-term periods in which we refit the spectral index are directly based on the 
periods identified via the Bayesian algorithm.
As discussed in \cite{HawcLightcurves}, the identification of these states is considered to be 
quite independent from the underlying spectral assumptions, since the lower energy limit for 
calculating the integrated photon fluxes was chosen in order to minimize differences in the fluxes. 
With a lower limit of 2 TeV for Mrk 421 and 3 TeV for Mrk 501, we expect only approximately 
$\pm5$\% variations in the photon fluxes for variations in the spectral parameters that cover the 
statistical and systematic uncertainty band of the average spectra.
For the comparisons in this contribution, on the other hand, we need to extrapolate the spectra to 
lower energies and integrate the differential fluxes above the threshold of FACT, 750\,GeV, in order 
to directly compare the measured values of the two instruments. It is thus important to account for 
the impact of varying intrinsic spectra that has been discussed for the sources under consideration, for 
example, in~\cite{Mrk421Krennrich2002,Albert2007Mrk501}.
Relying on these Bayesian segmentation, the increased statistics of many low-flux or a few high-flux transits included in each block made it possible to refit the power law index. We report results for the relevant blocks below with statistical uncertainties, though all spectral index values also have a systematic uncertainty of $\pm0.2$, as discussed in~\cite{HawcCrab}.
We then used these newly determined spectra to refit the normalization of each individual transit in the block, followed by a numerical integration above a threshold of 750\,GeV. The resulting flux values and correlations between the measurements with the two instruments are shown and discussed in the following.

\section{Results}

\subsection{Spectra}

Fitting the average 
intrinsic spectral model for Mrk 421 based on all 17 months of HAWC data showed that an intrinsic curvature, parametrized via an exponential cut-off at $7 \pm 2$ TeV, was preferred over a description with only a simple power law. Due to insufficient statistics for refitting two spectral parameters within each block, we decided to keep this cut-off fixed at 7 TeV and allowing only the power law index to vary. We will revisit the possibility of a varying intrinsic cut-off in the continuation of this study. We find that the indices vary between 
extreme values of 1.6 and 2.3 around an average of $2.1 \pm 0.1$. 

For Mrk 501, the average 
intrinsic spectrum, as measured with the combined HAWC data, is well described by a simple power law with index $2.0\pm 0.1$, showing no hint for intrinsic curvature in the energy range between 1.5 and 14 TeV. When refitting this index in each of the 14 
Bayesian Blocks, we find a clear indication of variability and index values range between 1.0 and 2.4 . 

As discussed above, these spectral parameterizations, with individual indices for each Bayesian block, are used to fit the normalization for the HAWC data of each transit. Integrating these spectra for each transit above the FACT threshold of 750\,GeV allows us to directly compare the HAWC and FACT photon flux values.

\subsection{Correlation of Nightly Averages}

\begin{figure}
\includegraphics[width=.48\textwidth]{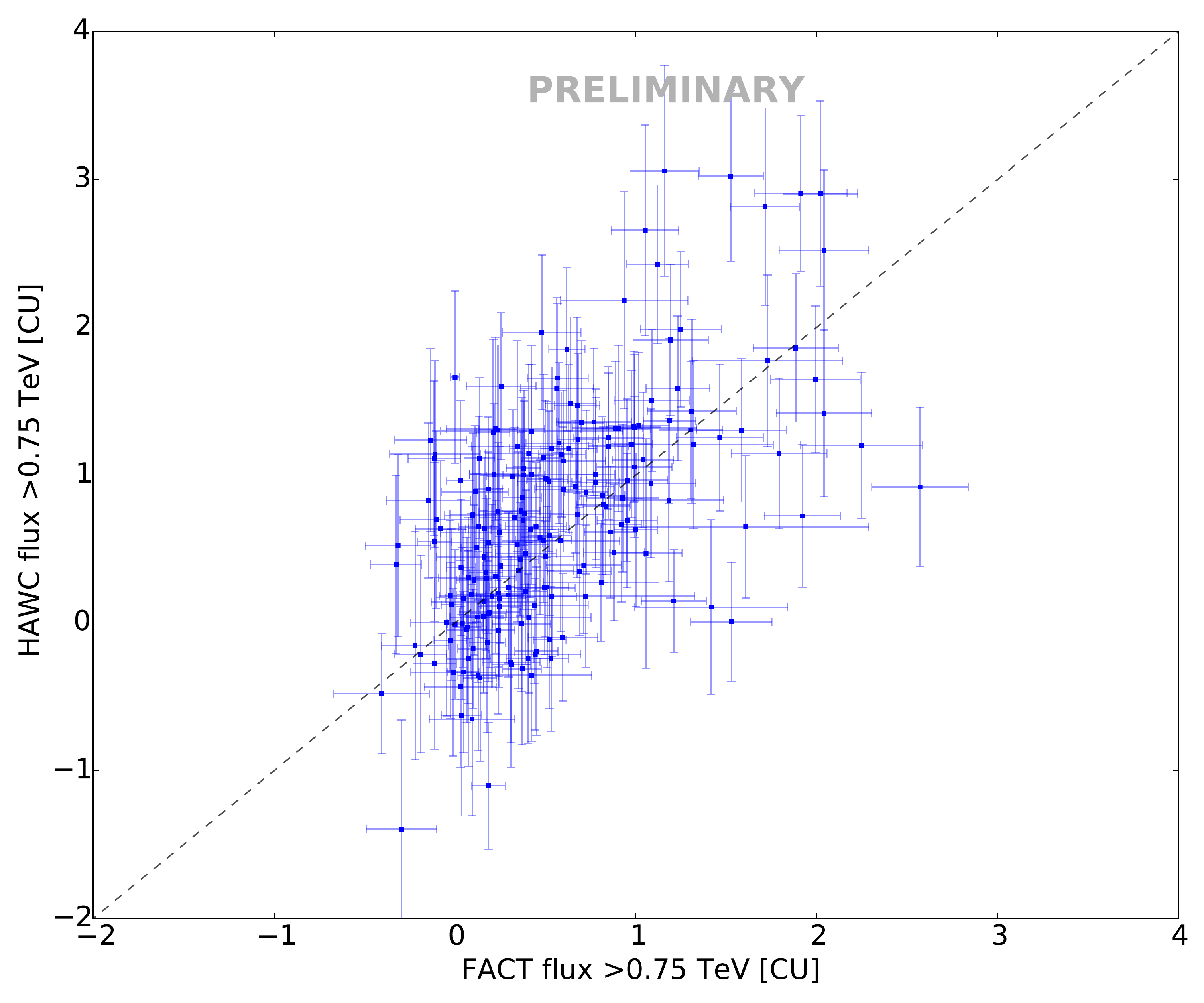}
\includegraphics[width=.48\textwidth]{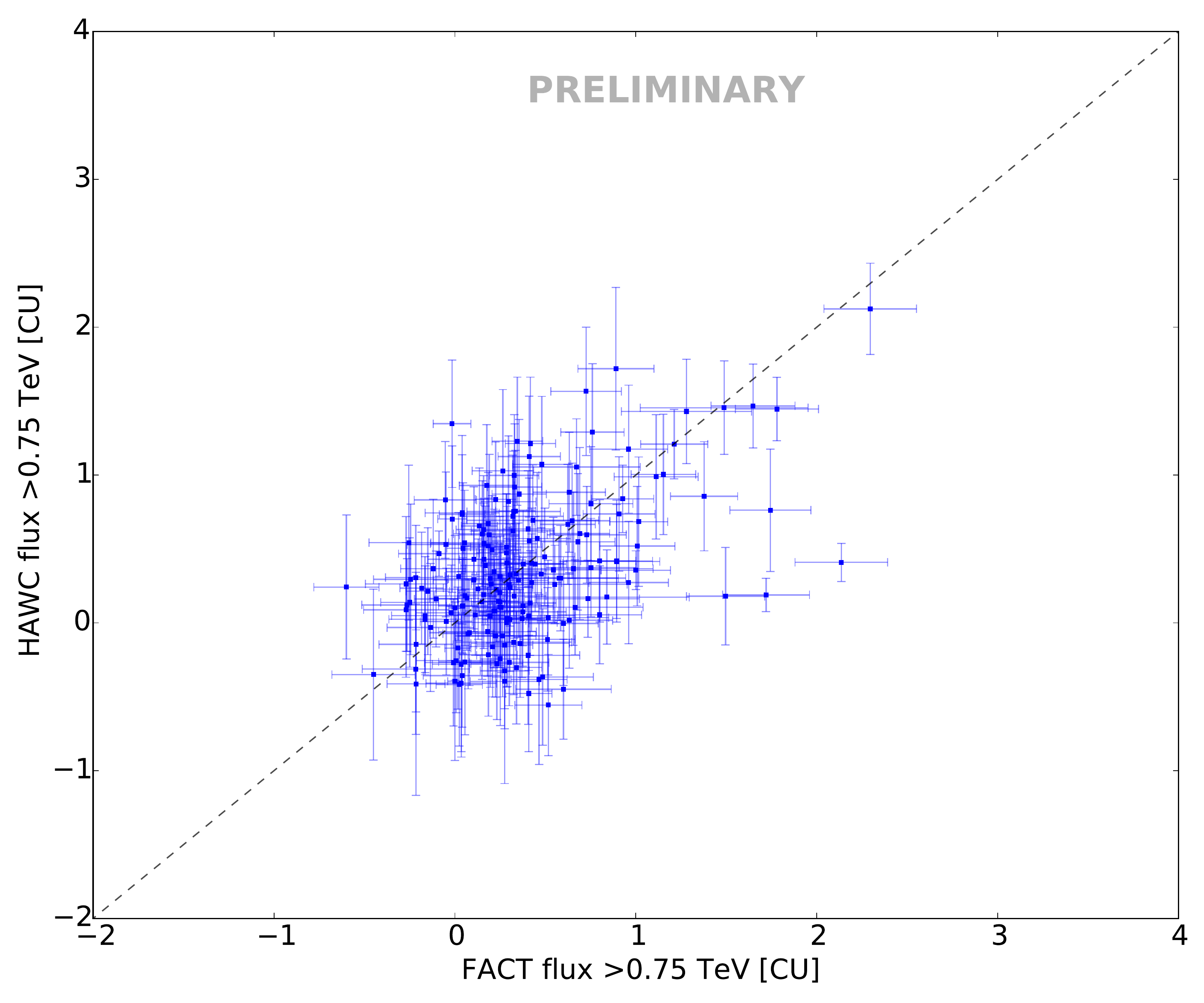}
\caption{Comparison of nightly average flux measurements from FACT and HAWC for 210 nights with coverage of Mrk 421 (left) and 200 nights for Mrk 501 (right).}
\label{fig:Corr}
\end{figure}

\subsubsection{Mrk 421}
On the left side of Figure~\ref{fig:Corr}, we show a scatter plot of flux values from Mrk\,421 for 210 nights 
for which measurements from both FACT and HAWC were available. The photon fluxes above a 
common threshold of 750\,GeV were converted to Crab Units (CU) based on direct Crab reference 
measurements in each experiment.
The Pearson correlation coefficient is $\rho=0.587$, with an approximate p-value 
of $2\cdot10^{-19}$ for occurring in case of no correlation of the underlying data. When we assume a perfect linear correlation, without offset, for a simulated data set of the same size and with values varying according to the same uncertainties then we find an expectation of $\langle \rho_{\mathrm{lin}} \rangle=0.697$ which is approximately three standard deviations larger than the observed value. We thus 
conclude that fluxes between FACT and HAWC are on average correlated but with a distribution that is probably not due to a simple 1:1 relationship. In view of the offset between 
the two locations and thus only minimal overlap in the time windows for data taking, this result seems to indicate that these average source fluxes change only relatively slowly over each $\sim 12$-hour period for most but not all of the nights.


\subsubsection{Mrk\,501}

The correlation plot for 200 measurements from nights covered by both experiments is shown in 
Fig~\ref{fig:Corr}. We obtain $\rho=0.475$ for the Pearson correlation coefficient with an approximate p-value of $8\cdot 10^{-12}$ for the null hypothesis of no correlation. Similar to the situation for Mrk\,421, the expected value for a linear correlation is approximately four standard deviations larger, $\langle \rho_{\mathrm{lin}} \rangle = 0.629$. 
This indicates, on average, a correlation between the fluxes measured with several hours offset by FACT and HAWC, but also shows evidence of various exceptions from this trend. In the plot, these exceptions are visible as outliers, in particular at least four with fluxes measured by FACT that are significantly ($>2$ standard deviations) higher than those observed by HAWC a few hours later. We will discuss a period that contains two of these nights in the following section.

\subsection{Study of Flaring Periods for Mrk\,501}

\begin{figure}
\includegraphics[width=.95\textwidth]{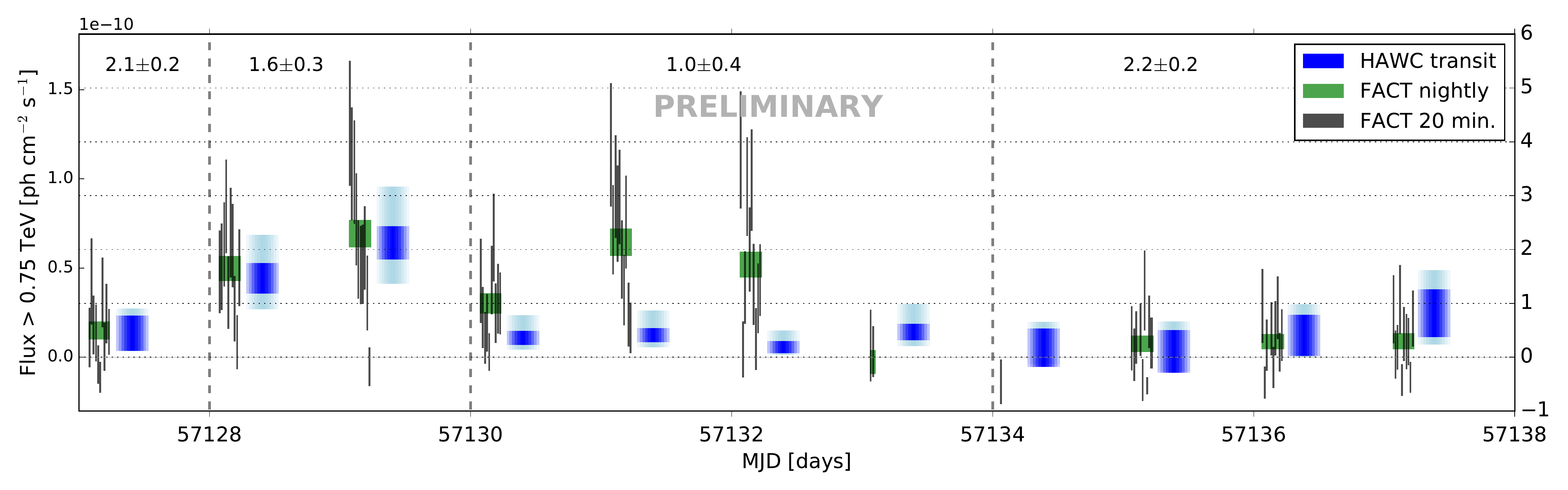}
\caption{Light curves for Mrk\,501 between 2015 April 15 and 25 with HAWC nightly average fluxes (blue), including statistical uncertainties from the normalization fit (inner, blue) and the fit of the index (outer, light blue), FACT nightly average fluxes (green), and FACT 20-minute measurements (gray). Also displayed are the spectral indices measured by HAWC (top row) and the edges of the Bayesian blocks (dashed lines).}
\label{fig:lcMrk501April2015}
\end{figure}

\begin{figure}
\includegraphics[width=.95\textwidth]{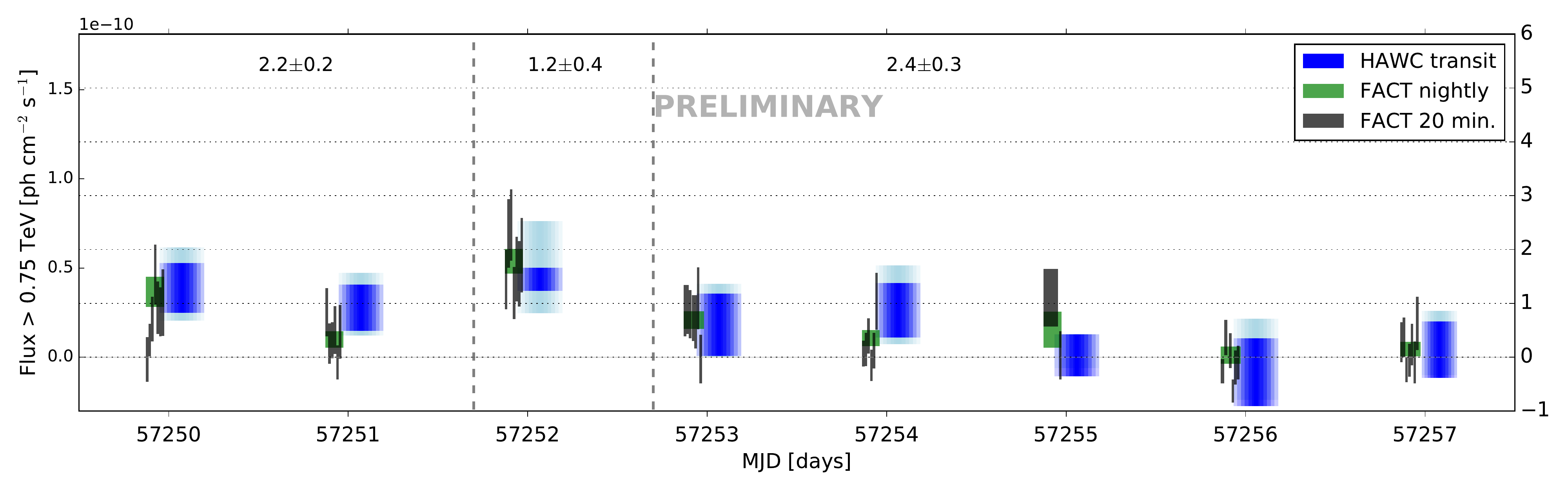}
\caption{Light curves for Mrk\,501 between 2015 August 16 and 23 with HAWC nightly average fluxes (blue), including statistical uncertainties from the normalization fit (inner, blue) and the fit of the index (outer, light blue), FACT nightly average fluxes (green), and FACT 20-minute measurements (dark gray). Also displayed are the spectral indices measured by HAWC (top row) and the edges of the Bayesian blocks (dashed lines).
}
\label{fig:lcMrk501August2015}
\end{figure}

In this section, we will focus on a case study of two flaring periods observed for Mrk\,501 and compare measurements from both FACT and HAWC in more detail. We reserve further discussion of other periods and results from Mrk\,421 for a future publication.
The light curve figures (Fig.~\ref{fig:lcMrk501April2015} and Fig.~\ref{fig:lcMrk501August2015} include HAWC transit averages (blue), FACT nightly averages (green), and FACT 20-minute measurements (dark gray). Horizontal width of each box indicates the time window of observation, with horizontal shading highlighting the varying sensitivity with zenith distance during a transit over HAWC. The vertical extend of each box shows the statistical flux uncertainty. In the case of HAWC, we also show in light blue the additional flux uncertainty stemming from the statistical uncertainty of the flux index fitted for each Bayesian block.

Two of the nights in which the FACT measurements show a significantly higher flux state than observed during the following transit over HACW occurred in April 2015, during adjacent MJDs 57131 and 57132. In Fig.~\ref{fig:lcMrk501April2015}, the combined light curve from both FACT and HAWC is shown starting from MJD 57120. 
The two nights belong to a 4-day period, MJD 57130 to 57133, that was identified in HAWC data as a distinct segment via the Bayesian blocks algorithm (marked with dashed lines in Fig.~\ref{fig:lcMrk501April2015}. The spectral fit for this period revealed an extremely hard intrinsic power law index of $1.0 \pm 0.4$. The larger fraction of multi-TeV photons arising from this spectral change leads to larger statistics in the most sensitive energy range for HAWC and thus to relatively small flux uncertainties for the HAWC measurement. 
In the night before this 4-day block, on MJD 57129, a very high flux state was observed, with similar nightly averages measured by both experiments. In the HAWC light curve, the period MJD 57128 to 57129 is identified via the Bayesian blocks as a distinct segment with a fitted power law index of $1.6 \pm 0.3$ that is softer than the that of the following period but harder than the average index of 2.1. 

The 20-min. resolution achieved by FACT highlights significant flux variations within a few hours for MJD 57129 as well as for the outlier nights MJD 57131 and 57132. For the latter two, the HAWC measurements are consistent with a low flux state after a downward trend during the FACT observations. On MJD 57129, on the other hand, the HAWC flux measurement seems to indicate that the source returned to a higher flux state after the flux decreased below that level a few hours earlier while FACT was observing.
In general, the period shown here includes some of the most dramatic changes in spectral indices observed with the HAWC analysis for Mrk\,501, and we also find significant variations of fluxes via the FACT measurements on time scales of less than one hour. Both phenomena seem to indicate extreme activity, leading to short-term fluctuations that explain differences in the fluxes observed by each instrument.

In Fig.~\ref{fig:lcMrk501August2015}, we show a period in August 2015 for which HAWC found another case of a hardened spectral index of $1.2 \pm 0.4$, in a segment identified via the Bayesian blocks algorithm that contains only the transit during MJD 57252. In this case, the FACT-measured nightly average obtained a few hours before the HAWC measurement is at a similar level as the HAWC flux point, while the 20-minute bins indicate some variations, though not as steep trends as in the example in Fig.~\ref{fig:lcMrk501April2015}. All of the other observations included in Fig.~\ref{fig:lcMrk501August2015} show relatively good correlation between the FACT and HAWC flux values, and the activity indicated by the spectral change seems to be contained within a very short time window, possibly not much longer than the 6 hours covered by HAWC. This transit has previously also been analyzed via the real-time flare monitor method in HAWC, tracing event rates on shorter time scales, see~\cite{Weisgarber2017}. The discussion in that reference indeed points to a short, few-hour flare, matching our observations presented here. 

\section{Conclusions}

In this contribution, we performed the first detailed comparison between TeV fluxes obtained with an Imaging Atmospheric Cherenkov Telescope, the FACT project, and a second generation, wide-field-of-view, water Cherenkov array, the HAWC Observatory. 
Even without a detailed discussion of instrument-specific systematic errors, we find generally very similar flux measurements above a common threshold. In particular, a clear correlations between nightly or 1-transit average fluxes of nights covered by both FACT and HAWC for the TeV blazars Mrk\,421 and Mrk\,501 is observed, indicating limited flux variations between the two observations. A few data points with significant differences between FACT and HAWC measurements point to interesting periods in which rapid variations of both the flux and the intrinsic spectral shape lead to changes in the source behavior within the 12-hour window that is near-continuously covered through the joint analysis. 

We will continue to study these flaring episodes of Mrk\,421 and Mrk\,501 in a joint analysis with several foreseeable improvements, such as inclusion of spectral measurements with FACT and subdividing the HAWC data into half transits in order to better trace short-term variability. A full tracing of flux and spectral changes will allow us to construct a joint, time-dependent spectral energy distribution that can be compared to models for the mechanisms that drive the rapid variability and TeV emission in these sources.

\acknowledgments

\noindent
HAWC Collaboration: \href{http://www.hawc-observatory.org/collaboration/icrc2017.php}{http://www.hawc-observatory.org/collaboration/icrc2017.php}\\ FACT Collaboration: \href{http://fact-project.org/collaboration/icrc2017_acknowledgements.html}{http://fact-project.org/collaboration/icrc2017\_acknowledgements.html}

\bibliographystyle{JHEP}
\bibliography{facthawc_icrc2017}

%

\end{document}